\documentclass[nofootinbib,prl,superscriptaddress,a4paper,twocolumn]{revtex4-1}
\usepackage{geometry}
\usepackage[english]{babel}
\usepackage[latin1]{inputenc}
\usepackage{hyperref}
\usepackage{amsmath, amssymb, amsfonts, mathrsfs}
\usepackage{graphicx}
\usepackage{enumitem}
\usepackage{xspace}

\newcommand*{\semnet}{{\normalsize S}{\scriptsize EM}{\normalsize N}{\scriptsize ET}\xspace}

\begin{document} 

\title{Predicting Research Trends with Semantic and Neural Networks\\with an application in Quantum Physics}

\author{Mario Krenn}
\email{mario.krenn@univie.ac.at}
\affiliation{Vienna Center for Quantum Science \& Technology (VCQ), Faculty of Physics, University of Vienna, Austria.}
\affiliation{Institute for Quantum Optics and Quantum Information (IQOQI), Austrian Academy of Sciences, Vienna, Austria.}
\affiliation{Department of Chemistry \& Computer Science, University of Toronto, Canada.}
\affiliation{Vector Institute for Artificial Intelligence, Toronto, Canada.}
\author{Anton Zeilinger}
\email{anton.zeilinger@univie.ac.at}
\affiliation{Vienna Center for Quantum Science \& Technology (VCQ), Faculty of Physics, University of Vienna, Austria.}
\affiliation{Institute for Quantum Optics and Quantum Information (IQOQI), Austrian Academy of Sciences, Vienna, Austria.}

\begin{abstract}
The vast and growing number of publications in all disciplines of science cannot be comprehended by a single human researcher. As a consequence, researchers have to specialize in narrow sub-disciplines, which makes it challenging to uncover scientific connections beyond the own field of research. Thus access to structured knowledge from a large corpus of publications could help pushing the frontiers of science. Here we demonstrate a method to build a semantic network from published scientific literature, which we call \semnet. We use \semnet to predict future trends in research and to inspire new, personalized and surprising seeds of ideas in science. We apply it in the discipline of quantum physics, which has seen an unprecedented growth of activity in recent years. In \semnet, scientific knowledge is represented as an evolving network using the content of 750,000 scientific papers published since 1919. The nodes of the network correspond to physical concepts, and links between two nodes are drawn when two physical concepts are concurrently studied in research articles. We identify influential and prize-winning research topics from the past inside \semnet thus confirm that it stores useful semantic knowledge. We train a deep neural network using states of \semnet of the past, to predict future developments in quantum physics research, and confirm high quality predictions using historic data. With the neural network and theoretical network tools we are able to suggest new, personalized, out-of-the-box ideas, by identifying pairs of concepts which have unique and extremal semantic network properties. Finally, we consider possible future developments and implications of our findings.
\end{abstract}
\date{\today}
\maketitle 

\section{Introduction}
A computer algorithm with access to a large corpus of published scientific research could potentially make genuinely new contributions to science. With such a body of knowledge, the algorithm could derive new scientific insights that are unknown to human researchers and note contradictions within existing scientific knowledge \cite{evans2011advancing,you2015darpa}. This level of automation of science is more in the realm of science-fiction than reality at present. However, algorithms with access to and the capability of extracting semantic knowledge from the scientific literature can be employed in manifold ways to assist scientists and thereby augment scientific progress. As an example, the evaluation of whether an idea is novel or surprising depends crucially on already-existing knowledge. Thus a computer algorithm with the capability to propose new, useful ideas or potential avenues of research will necessarily require access to published scientific literature - which forms at least partially the body of human knowledge in a scientific field.

\begin{figure}[!b]
\centering
\includegraphics[width=0.50\textwidth]{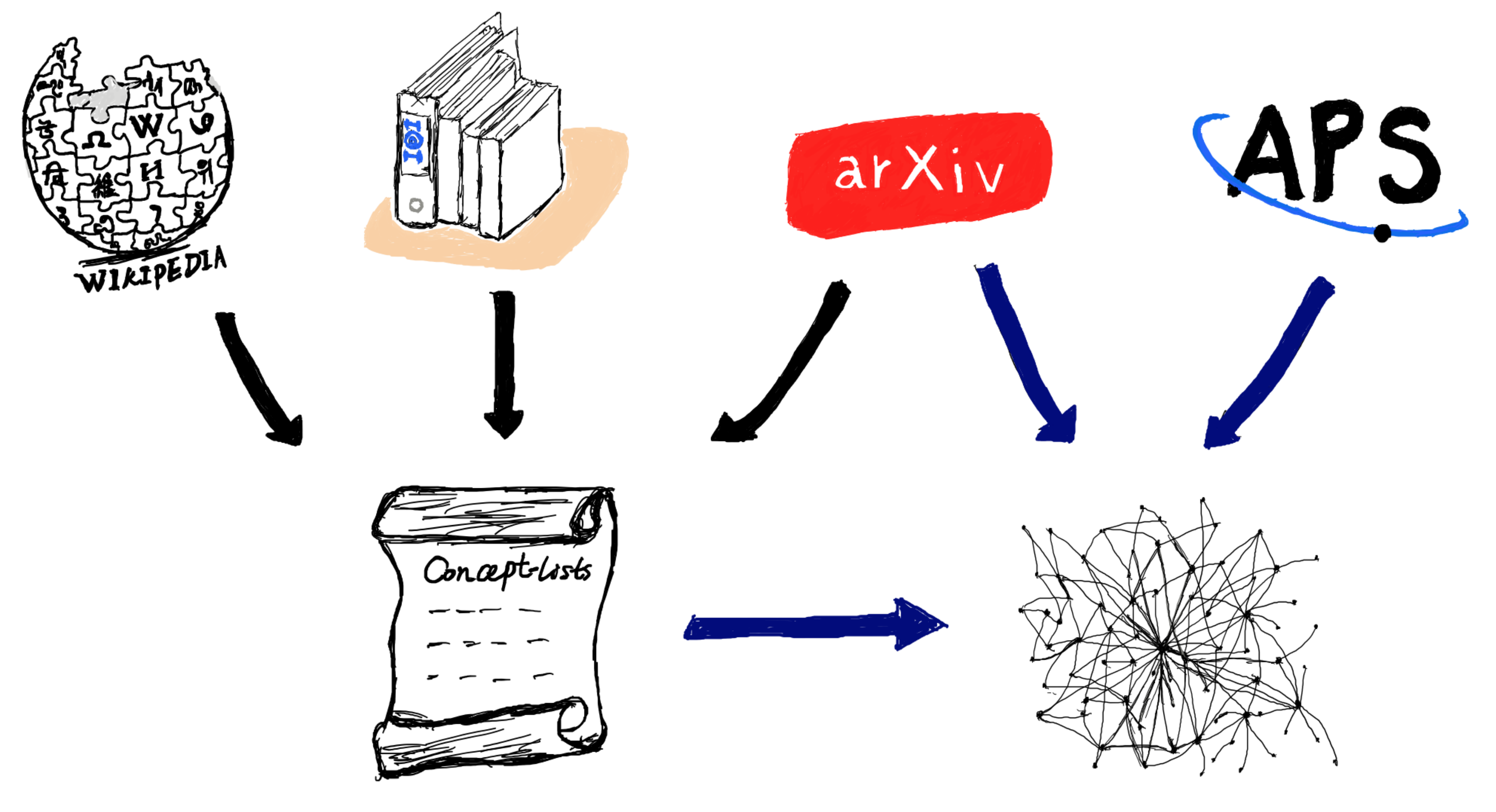}
\caption{Creating a semantic network for quantum physics (\semnet). The nodes represent quantum physical concepts, and the edges (connections between nodes) indicate how frequently two concepts are investigated jointly in the scientific literature. The concept list is created using human-made lists (from Wikipedia categories and quantum physics books) and automatically generated lists using natural language processing tools on 100.000 quantum physics articles from the online preprint repository arXiv (this is indicated by black arrows). An edge between two concepts is drawn when both concepts appear in the abstract of a scientific paper (indicated by blue arrows). The scientific database consists of 750.000 physics papers, 100.000 from arXiv and 650.000 papers published by the American Physical Society (APS) since 1919.}
\label{fig:CreateNetwork}
\end{figure}

Knowledge can be portrayed using semantic networks that represent semantic relations between concepts in a network \cite{lehmann1992semantic}. Over the last few years, significant results have been obtained by automatically analyzing the large corpus of scientific literature \cite{evans2011metaknowledge,zeng2017science,fortunato2018science}, including the development of semantic networks in several scientific disciplines. 

In biochemistry, a semantic network has been built using a well-defined list of molecule names (which correspond to the nodes of the network) and forming edges when two components co-appeare in the abstract of a scientific paper. The network was derived from millions of papers published over 30 years, and the authors identify a more efficient, collective strategy to explore the knowledge network of biochemistry \cite{foster2015tradition, rzhetsky2015choosing}. In \cite{iacopini2018network}, a semantic network was created using 100.000 papers from astronomy, ecology, economy and mathematics. The nodes represent ideas or concepts (generated through automated generation of key-concepts in large bodies of texts \cite{milojevic2015quantifying}). The authors used the network to draw connections between human innovation process and random walks. In the field of neuroscience, semantic networks have been used to map the landscape of the field \cite{beam2014mapping, dworkin2018landscape}. Papers from the interdisciplinary journal PNAS have been used to investigate sociological properties such as inter-disciplinary research \cite{dworkin2019emergent}.

\begin{figure}[!t]
\centering
\includegraphics[width=0.475\textwidth]{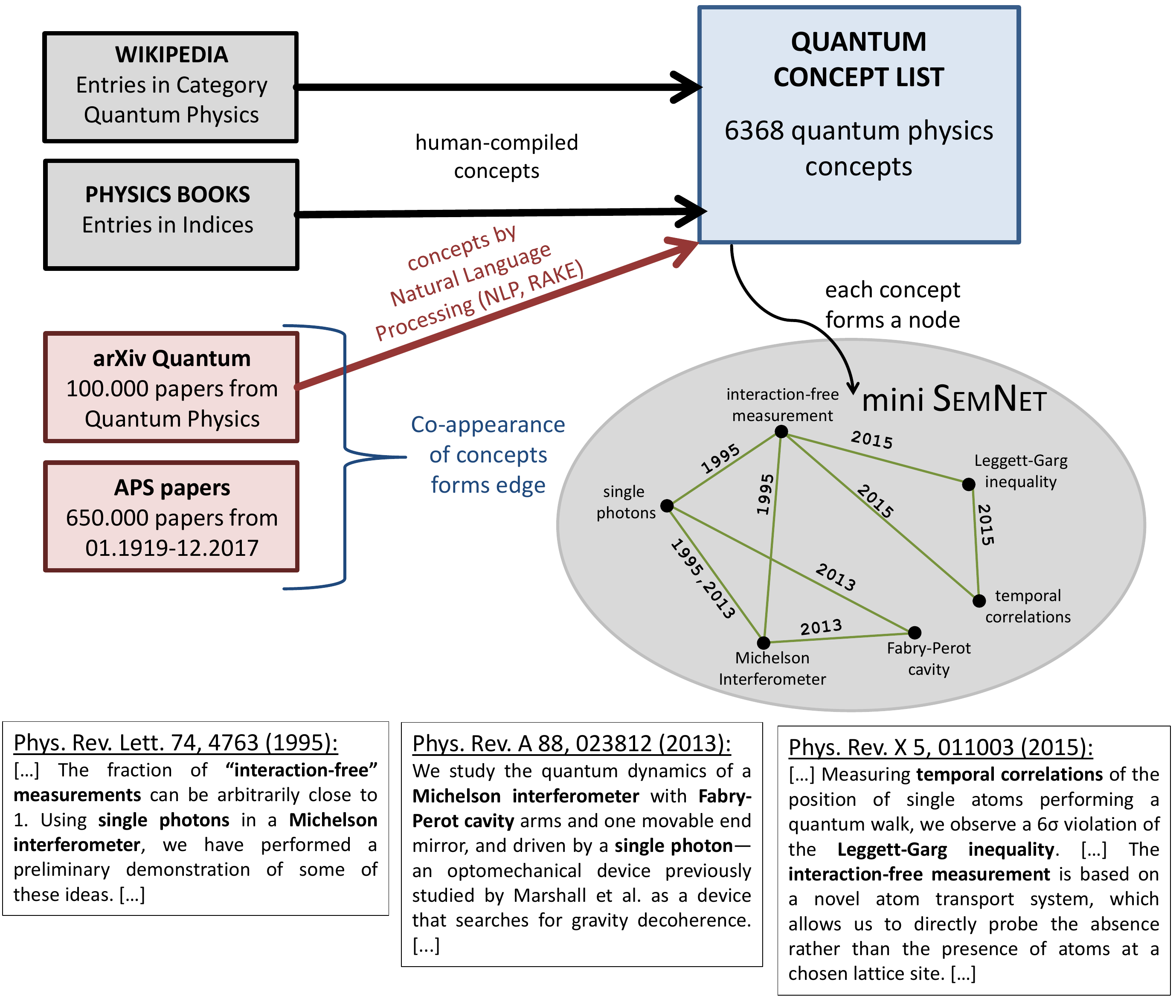}
\caption{Diagrammatic inner working of \semnet. Human-generated concept lists (from Wikipedia and books) are combined with automatically generated lists (with natural language processing, using RAKE on 100.000 arXiv articles) to generate a list of quantum physics concepts. Each concept forms a link in a semantic network. The edges are formed when two concepts co-appeare in a title or abstract of any of the 750.000 papers (from arXiv and APS). A mini-version of \semnet is shown, using parts of three articles from APS. Edges carry temporal information of their formation year, which leads to an evolution of the semantic network \semnet over time.}
\label{fig:CreateNetwork2}
\end{figure}

Here, we show how to build and use a semantic network for quantum physics, which we call \semnet. It is built from 750.000 scientific papers in physics published  since 1919. In the network we identify a number of historic award-winning concepts, indicating that \semnet carries useful semantic knowledge. The evolution of such a large network allows us to use an artificial neural network for predicting research concepts that scientists will investigate in the next five years. Finally, we demonstrate the power of \semnet to suggest personalized, novel and unique directions for future research \footnote{Code and details: \url{https://github.com/MarioKrenn6240/SEMNET}}.  

Our work differs in several aspects from previous semantic networks created from scientific literature. First, we use machine learning to draw conclusions from earlier states to \semnet's future state, which enables us to make predictions about the future research trends of the discipline. Second, we use network theoretical tools and machine learning to identify pairs of concepts with exceptional network properties. Those concept combinations can be restricted to the research interest of a specific scientist. This ability allows us to not only predict but also suggest uninvestigated concept pairs which human scientifists might not have identified because they are out of the own sub-field, but which have properties that indicate an exceptional relation. They could be a seed of a new, out-of-the-box idea. Third, we apply \semnet to quantum physics, which has seen an enormouse growth during the last decade due to the potential transformative technologies. The growth can be seen in the establishment of several high-quality journals for quantum research (such as Quantum, npj Quantum Information, IOP's Quantum Science \& Technology) and multi-billion dollar fundings from governments and strong involvement of private companies and startups worldwide. The growth rate leads to enormous increase in scientific results and publications, which are difficult to follow for individual researchers -- thus quantum physics is an ideal test-bed for \semnet.

\section{Semantic Network of Quantum Physics}
\begin{figure*}[!t]
\centering
\includegraphics[width=0.94\textwidth]{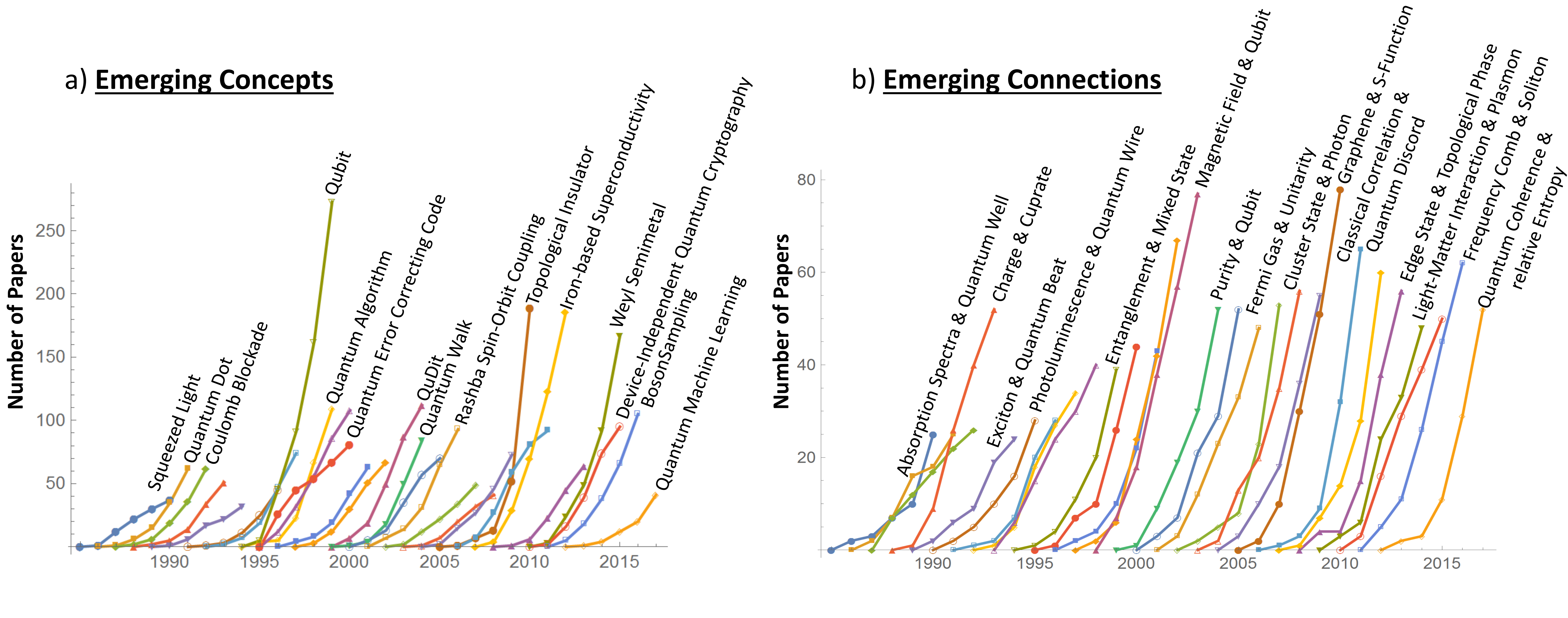}
\caption{The evolution of quantum physics research observed using \semnet, reflected in the change in number of articles that contain a concept or concept pair per year from 1987 to 2017. (a) Newly-emerged concepts and their growth in popularity over a five-year period after emergence. Shown are the strongest growing concepts of a five-year period, which have not been mentioned before that period. (b) Newly-connected pairs of concepts that become strongly influential in the scientific community in a five-year period. Shown are the strongest growing connections of concept pairs that already existed before the connection was drawn, which have not been connected before that period. Many emergent concepts and connections can be related to important discoveries and understandings in quantum science.}
\label{fig:Emergers}
\end{figure*} 

A semantic network, or knowledge network, represents relations between concepts in the form of a network. Now we describe in more detail how the network is built, especially how the concept list is generated and how links are formed. A schematic illustration can be seen in Figure \ref{fig:CreateNetwork}, more details in Figure \ref{fig:CreateNetwork2}.
\subsection{Creation of the concept list}
We generate the concept list via two independent methods. First, we use human-made lists of physical concepts. These concepts are compiled from the indices of 13 quantum physics books (which were available to us in a digital form), as well as titles of Wikipedia articles that are linked in a quantum physics category. This human-made collection contains approximately 5000 entries physical concepts.

We extend the human-generated list with an automatically generated list of physical concepts. For this, we apply a natural language processing tool called RAKE (Rapid Automatic Keyword Extraction) \cite{rose2010automatic} to the titles and abstracts of approximately 100.000 articles published in quantum physics categories on the arXiv preprint server, which we chose to optimize the list for current research topics in quantum physics. RAKE is based on statistical text analysis, and can automatically find relevant keywords in texts. We combine the human- and machine-generated lists of concepts and further optimize them to delete incorrectly identified concepts (which were introduced by imperfections of the statistical analysis of RAKE) and names of people (which are not concepts), merge synonyms and normalize for the singular and plural of the same concept. Ultimately, this yields a list of 6,300 terms. As an example, five randomly chosen examples are \textit{three level system}, \textit{photon antibunching}, \textit{chemical shift}, \textit{neutron radiation} and \textit{unconditionally secure quantum bit commitment}. Each of these quantum physics concepts is a node in \semnet.

\subsection{Creation of the network} 
To form connections between different quantum physics concepts, we use 100.000 articles of quantum physics categories on arXiv, and the dataset of all 650,000 articles ever published by the APS. We chose these two data sources because the APS database contains peer-reviewed physics papers from the last 100 years (allowing for investigation of long-term trends), while the arXiv database contains specific quantum physics papers, allowing for more precise coverage of the quantum physics research trends.

\begin{figure*}[!t]
\centering
\includegraphics[width=0.9\textwidth]{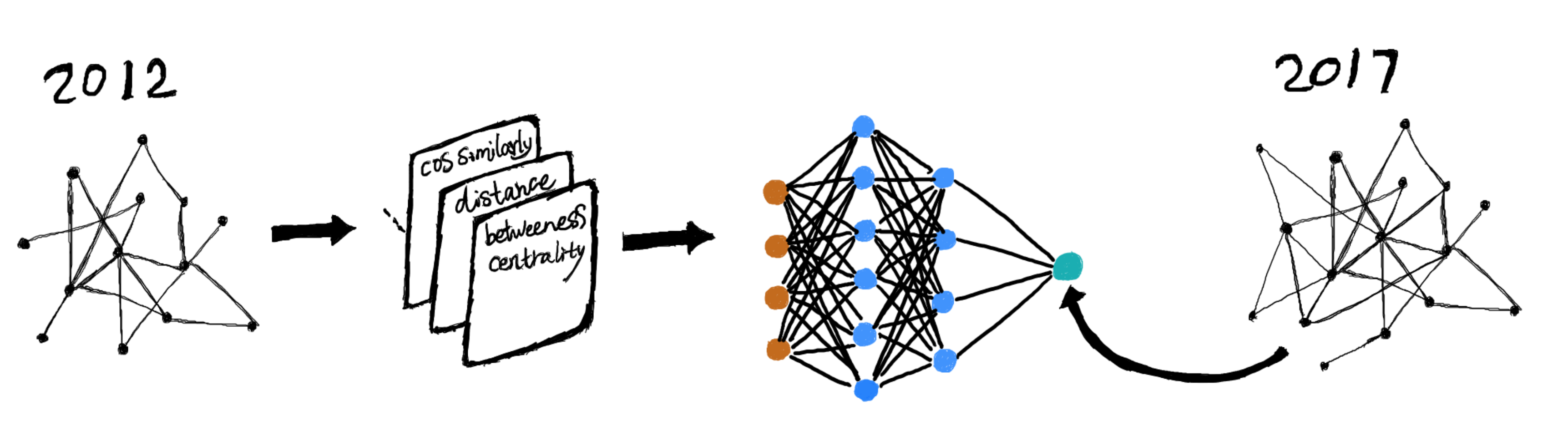}
\caption{Artificial Neural Network for predicting the future of quantum physics research, using the evolution of the semantic network \semnet. For each unconnected pair of concepts at a specific year, we derive a vector of 17 network properties (such as distance or cosine similarity). In the training phase, we input these network properties into an artificial neural network, and ask the question whether they will be connected 5 years later. \semnet of 2017 is used for supervision. After training, we can apply the neural network to \semnet of 2017, and ask what will have happend until the year 2022.}
\label{fig:TrainNN}
\end{figure*} 

Whenever two concepts occur together in a title or an abstract of an article, we interpret that as a semantic connection between these concepts, and add a unique link between the two corresponding nodes in the network. Relations between two concepts can take many forms. Concepts may be put together for example when mathematical tool (such as \textit{Schmidt rank}) is used to investigate a specific quantum system (such as \textit{vector beam} or \textit{exciton polariton}), or when insights from a specific technique (such as \textit{lasing without inversion} or \textit{rabi oscillation}) lead to conclusions about another property (such as \textit{transport property} or \textit{atom transition frequency}) or when fundamental ideas (such as \textit{quantum decoherence} or \textit{quantum energy teleportation}) are studied in the context of foundational experiments (such as \textit{delayed choice experiment} or \textit{Mermin inequality}). While this method clearly cannot represent all quantum physics knowledge, it represents elements of its semantic structure, which we demonstrate in what follows. 

The resulting network \semnet has 6368 vertices with more than 1.7 million edges (drawn from more than 15 million concept pairs pulled from 750.000 physics articles), using physics articles from 1919 to december 2017.

\section{Results}
\subsection{Past quantum physics trends}

First, we use the evolution of the semantic network to identify impactful emerging fields of research in the past. We define \textit{emerging fields} as either concepts or concept pairs which have grown significantly after they have been introduced or connected for the first time, over periods of five years.

Figure \ref{fig:Emergers}a shows the quantum physics topics that have grown the fastest (in terms of numbers of papers in which they have been mentioned) after their emergence, from the years 1987 to 2017. Figure \ref{fig:Emergers}b shows, for each year, which two-concept combinations have grown the fastest in the first five years after they have been first connected. In Figure \ref{fig:Emergers}, many of the emerging fields clearly correspond to important discoveries, advances in understanding and shifts of thought within quantum science research. One of the fastest growing concepts is \textit{Qubit}, which emerged in 1995 (first in april in a Phys.Rev.A paper by Schumacher \cite{schumacher1995quantum}, then in arXiv preprints by Chuang\&Yamamoto \cite{chuang1995simple} and by Knill \cite{knill1995approximation, knill1995bounds}). Qubits are the basic units of quantum information -- generalizing classical bits to coherent quantum superpositions, and connect quantum mechanics and information science. The emergence of the qubit can be interpreted as the start of the discipline of \textit{quantum information science}. Enormous growth is seen for topics connected to graphene, starting in 2005, the discoverers of which were awarded the 2010 Nobel Prize in Physics. Interesting, graphene itself was mentioned (in our data collection) already back in the early 1990s in Phys.Rev.B papers \cite{bayot1990two, di1991magnetic, moreh1991effective}, when it was not a strongly emergent concept itself. Strong growth in research into topological materials can be observed from approximately 2008; the Nobel Prize in Physics was subsequently awarded in this area in 2016. Aaronson's and Arkhipov's approach to achieving \textit{quantum supremacy} \cite{harrow2017quantum} using linear photonic networks, termed BosonSampling \cite{aaronson2011computational}, achieved considerable attention (with more than 600 citations since its introduction in 2011, and considerable experimental efforts into this directions). Since 2012, the application of machine learning to quantum physics has become a prominent and diverse topic of research, that falls under the umbrella of \textit{quantum machine learning} (recently summarized in two prominent reviews \cite{biamonte2017quantum,dunjko2018machine}, and also observable by the foundation of a novel high-quality journal for this topic, Springer Quantum Machine Intelligence). These findings confirm that \semnet contains useful semantic information.
 
\subsection{Predictive ability of the \semnet}   
Having used \semnet to study past quantum trends, we investigate its ability to provide projections of knowledge developments in the future. This essential question in network science is called \textit{link-prediction problem}, and asks which new link will be formed between unconnected vertices of the network in the future given the current state of the network (for a detailed investigation of the link-prediction problem in network theory, see \cite{liben2007link}). We apply this problem in the context of semantic networks which are generated from published scientific literature. In the present case looking at the field of quantum physics, we ask which two concepts that have not yet been studied together might be investigated together in a scientific article over the next five years. To answer this question, we use an artificial neural network, with four fully connected layers (two hidden layers). The structure of the neural network and its training is shown in Figure \ref{fig:TrainNN}. Its task is to rank all unconnected pairs of concepts (roughly 5\% of all edges have been drawn by the end of 2017), starting with the pair that is most likely to be connected five years, up to the pair that most likely stays unconnected. Ultimately we want to apply the neural network to the current \semnet and predict the future trends. To validate its quality, we first input to the neural network past states of \semnet (for example, containing data only up to 2002), and train it to predict new links by 2007. After the training, we apply this network on 2007 data and validate its quality for data of the year 2012 (which it has never seen before). 
 
The semantic network is very large (consisting of 6368$\times$6368 entries for each year, which are the number of possible connections between the 6368 quantum physics concepts, compared to 28$\times$28 pixels for the famous MNIST dataset of handwritten images, and 256$\times$256 pixels for ImageNet \cite{lecun2015deep}), and involves combinatorial, graph-based information which are more structured than images (see for example \cite{wu2019comprehensive}). For that reason, it is an unsuitable direct input to the neural network. Instead, we compute semantic network properties for each pair of concepts. For each pair of concepts $(c_i,c_j)$ that are unconnected in \semnet, we calculate 17 network properties $p_{i,j}=(p_{i,j}^1, p_{i,j}^2, \dots, p_{i,j}^{17})$ where $p_{i,j}^k \in \mathbb{R}$. Here, $p_{i,j}^1$ and $p_{i,j}^2$ are the degrees of concept $c_i$ and $c_j$, and $p_{i,j}^3$ and $p_{i,j}^4$ are the numbers of papers in which they are mentioned. While these four properties are purely local, $p_{i,j}^5$ is the cosine-similarity between the two concepts, which corresponds to the number of common neighbors. A cosine similarity of one indicates that the terms might be synonyms. The next nine properties indicate the number of paths with lengths of two, three and four between the physics concepts in the current and previous two years. These properties allow us to draw conclusions from the evolution over time of various topics as tracked by \semnet. The choice to use large path lengths as one of the properties is strengthened by a very recent observation that the paths of length 3 (L3) are crucial for link prediction tasks in a network for protein interactions \cite{kovacs2019network}. Finally, the last three properties correspond to three different measures of distance between the two concepts. More details can be seen in the SI.  

\begin{figure}[!t]
\centering
\includegraphics[width=0.4\textwidth]{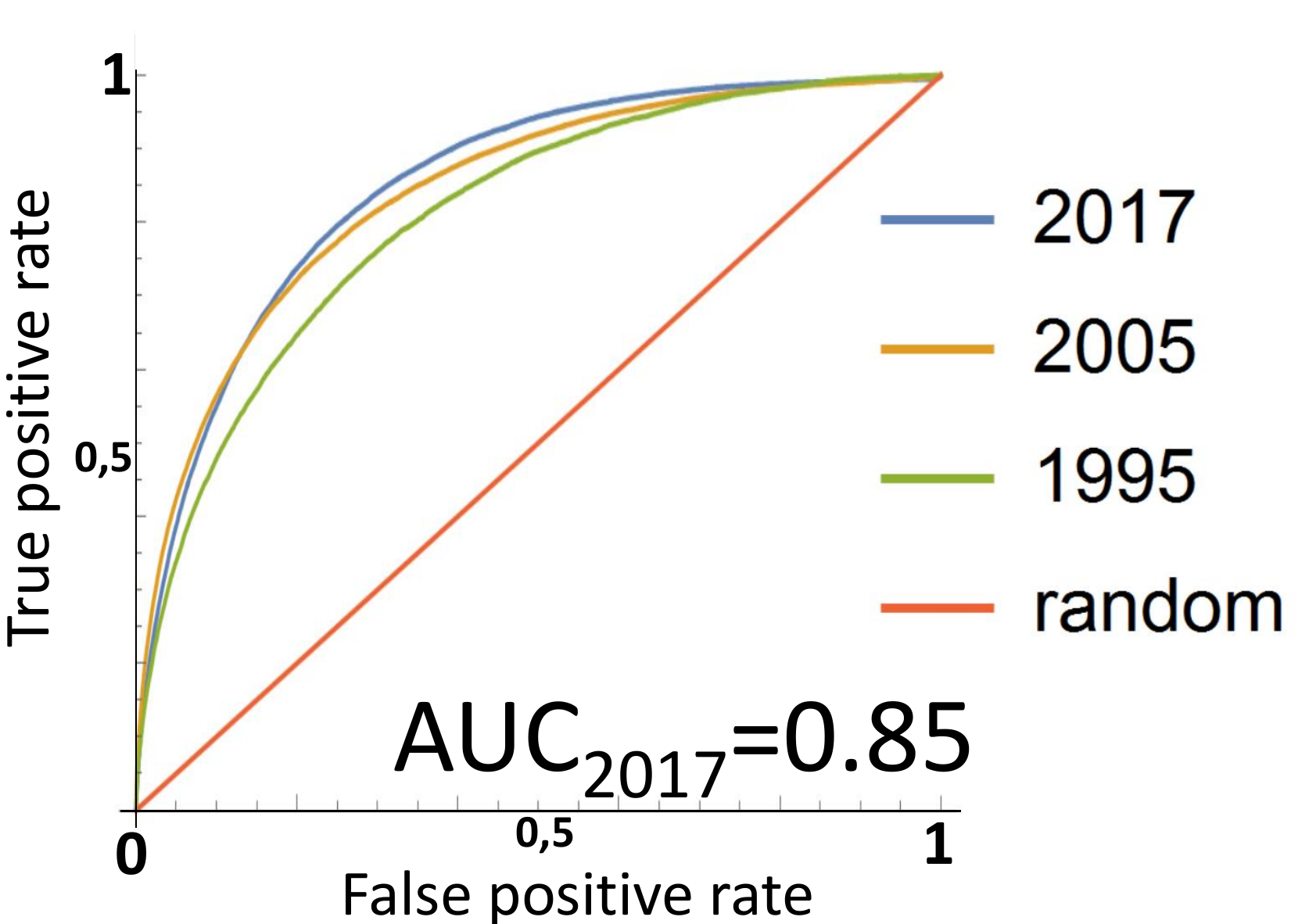}
\caption{Quantifying the prediction quality of the neural network regarding whether unconnected pairs will be connected within 5 years, using a receiver operating characteristic (ROC) curve. The y-axis shows the true-positive (TP) rate (rate of pairs that have been correctly identified to be connected within 5 years). The x-axis shows the false positive (FP) rate of predictions -- concept pairs that have falsely been predicted to be connected. We restrict ourselves to concept pairs which share less than 20\% of their neighbors, to prevent predictions of terms with similar semantical meaning. A perfect neural network would have TP = 1 while FP = 0. A network that classifies 50\% of true instances correctly, and misclassifies 10\% false instance as true would have TP = 0.5 and FP = 0.1. A random classifier is incorrect half the time and thus lies along the diagonal. The area under the curve (AUC) for a perfect neural network is 1, while for random predictions, it is AUC = 0.5. The AUC can be interpreted as the probability that the neural network will rank a randomly chosen true instance higher than a randomly-chosen negative instance \cite{fawcett2004roc}. The ROC validation curves for 1995, 2005 and 2017 (trained with \semnet using data from only 1990, 2000 and 2012 and earlier, respectively) are consistently and significantly non-random, with AUC$_{2017}$ = 0.85. These results show that the neural network can learn to predict future research interests in quantum physics, based on historical information to a high accuracy.}
\label{fig:ROC}
\end{figure}

We explain these properties on a concrete pair of concepts, \textit{interaction-free measurement} and \textit{Leggett-Garg inequality}. (We chose the example randomly, from unconnected concepts that had been mentioned individually more than 30 times.) The concept $c_{2526}$ represents "interaction-free measurement which is mentioned in 60 abstracts and has 135 connections to other concepts by 2012. The concept $c_{2819}$ represents the "Leggett-Garg inequality", which occurs in 33 abstracts and has 141 connections to other concepts by the end of 2012. These two concepts were not connected in \semnet as of 2012, therefore, the 15th property, their network distance, is $p_{2526,2819}^{15}=2$ (neighbors have a distance of one, in other words, there is a direct path connecting them of length one). In 2012, the two concepts have a cosine-similarity $p_{2526,2819}^{5}=0.228$, meaning that 22.8\% of their neighbors are shared. Two years later, in 2014 an article on arXiv mentioned both of these concepts in the abstract and the work was later published \cite{robens2015ideal} and featured \cite{knee2015quantum} in the high-impact journal \textit{Physical Review X}, achieving approximately 100 citations within four years. This example indicates that drawing first connections between concepts can lead to significant scientific insights.
 
The 17 properties for each unconnected concept pair in \semnet are used by the neural network to estimate which pairs of quantum physics concepts are likely to be connected within 5 years and which are not. 

To quantify the quality of the predictions, we employ a commonly-used technique called the \textit{receiver operating characteristic (ROC)} curve \cite{fawcett2004roc}. For this, the neural network is used to classify unconnected nodes into two sets: one set that is connected after five years, and a set that is non-connected. Figure 4 shows a significant ability to predict connections between pairs of topics -- even through we restrict ourselves to pairs that share less than 20\% of their neighbors (to prevent predictions of concepts which have similar meaning). This indicates that even research that draws new connections between concepts, can be predicted with high quality.

\section{Proposing future research topics} 

\begin{figure}[!t]
\centering
\includegraphics[width=0.4\textwidth]{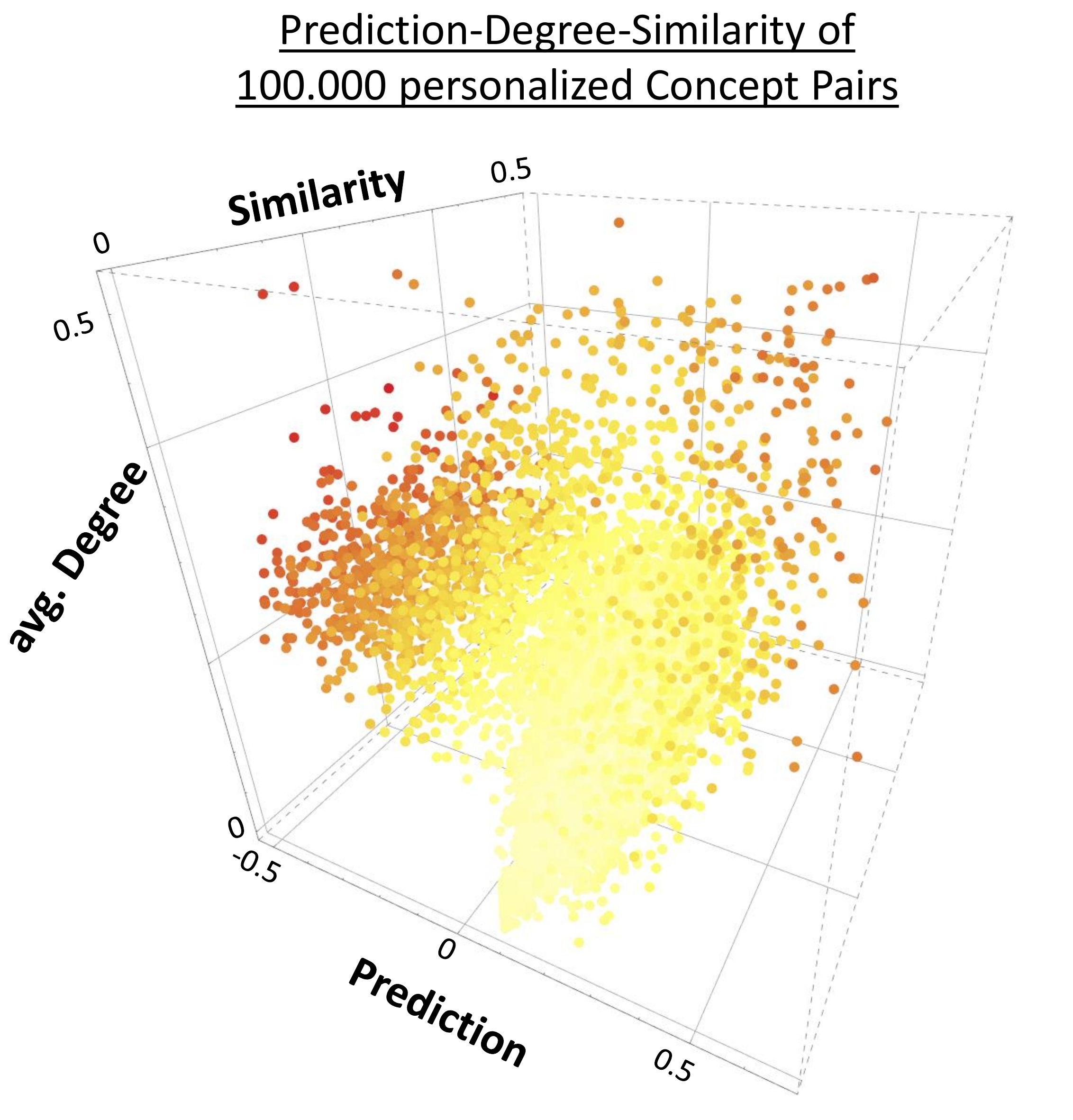}
\caption{Personalized prediction of topic pairs that could form future research directions for a given scientist. Each dot represents one unconnected pair of physical concepts. The concepts in use are filtered by a scientist's previous research agenda (see main text). The dot is placed in a three-dimensional space, which is proscribed by the properties of \semnet and the predictions of the neural network. One axis is the neural network predictions of whether two unconnected points will be connected in 2022 (the predictions -0.5 stand for very unlikely, 0.5 is very likely). The y-axis represents the average (normalized) degree of the pair (the concept with the highest degree in the complete network has a degree of 1). The z-axis is the cos-similarity, which is the ratio of shared neighbors in the networks of the two concepts. The color of the dots represents the distance from the most common, average point in this space -- darker dots are further away from the average. Outliers represent pairs of concepts with a unique network property, which make them ideal candidate suggestions.}
\label{fig:outlier}
\end{figure}
Next, we attempt to use \semnet and the artificial neural network to suggest new, potentially fruitful research directions in quantum physics. While it is interesting and useful to understand future trends, it potentially cannot by itself lead to surprising or out-of-the-box ideas (otherwise they would not be predictable). Therefore, we extend our previous approach with network theoretic tools, to identify concept pairs with exceptional network-theoretic properties. Furthermore. Since science is conducted by (groups of) individual scientists, suggestions for proposed new research directions need to be personalized (otherwise, we would obtain suggestions for topics in which nobody is an expert in -- which may be potentially interesting but limited in applicability).

How do we obtain suggestions for an individual scientist? What we find interesting and surprising strongly depends on what we already know. To gauge that, we need to investigate a given scientist's previously- published body of research papers and extract a list of concepts (from the concept list generated before) that define that person's personal research agenda(s). We define key concepts as concepts investigated over-proportionally often by the scientist, compared to the relative frequency of that concept in all 750.000 papers. Each concept $c_i$ in the papers authored by the scientist has a probability $p_{\textnormal{scientist}}(c_i)$ that we calculate by the the number of occurrences of the concept $N(c_i)$ divided by the sum of occurrences of all concepts, which is $p_{\textnormal{scientist}}(c_i)=\frac{N(c_i)}{\sum_j N(c_j)}$. Each concept also has a probability of occuring in all 750.000 papers that we use, written as $p_{\textnormal{total}}(c_i)=\frac{M(c_i)}{\sum_j M(c_j)}$, where $M(c_i)$ is the number of occurrences of the concept $c_i$ in all 750.000 articles. The ratio $r_{\textnormal{scientist}}(c_i)=\frac{p_{\textnormal{scientist}}(c_i)}{p_{\textnormal{total}}(c_i)}$ indicates the research agenda of the scientist. A value of $r_{\textnormal{scientist}}(c_i)>1$ shows that the scientist investigates the concept $c_i$ overproportionally often.
 
Our approach is to identify personalized suggestions of pairs of concepts that have never been connected. The concepts with $r_{\textnormal{scientist}}(c_i)>1$ value are paired with all of the other 6.368 concepts. This translates to a list of potentially 100.000s of possible topic pairs. For further usability, we introduce a way to sort the candidate suggestions. Suggestions can be sorted by identifying concept pairs with unique and unusual properties. For each pair of concepts, we have already calculated 18 different network properties: 17 properties which have been used by the neural network for generating predictions, and the prediction value itself. Together, these properties define a multi-dimensional space in which the location of each concept pair depends on its network properties.

To identify unusual and unique concept pairs, we search for outliers in this high-dimensional space. An outlier indicates a pair of concepts that is uniquely located in the space, and thus has unique properties in the semantic \semnet network. We can visualize, for an anonymous example scientist, a 3-dimensional projection of the high-dimensional space in Fig. \ref{fig:outlier}. There, every dot corresponds to a concept pair which is located according to its network properties. Outliers can be identified by the darkness of their color.

A few suggestion from \semnet, for the example scientist: Some of the highest predicted pairs (from Top10) are \textit{orbital angular momentum} \& \textit{magnetic skyrmion}, \textit{spin orbit coupling} \& \textit{quantum sensing} or \textit{dicke model} \& \textit{cloning}, filtered for highly predicted, uncommon pairs (cosine similarity $<$ 0.03; from Top10): \textit{topos theory} \& \textit{cyclic operation}, \textit{critical exponent} \& \textit{reed muller code}, \textit{quantum key distribution} \& \textit{adhm construction}. Unrestricted concept lists (normalized concept degree $<$ 0.1; from Top10): \textit{atom cavity system} \& \textit{mode volume}, \textit{entanglement of formation} \& \textit{multiqubit state}, \textit{neutrino oscillation} \& \textit{dark photon}. For more examples, see SI. 
 
\section{Outlook} 
\textbf{Machine Learning} -- Graph-based machine learning models, which have been studied in recent years, could improve prediction qualities in the link-prediction task, for example see \cite{li2015gated, niepert2016learning, wu2019comprehensive}. Furthermore, as \semnet represents a time evolution of quantum physics' semantic network, applying efficient tools for handling time-dependent data, such as a \textit{long short-term memory} \cite{hochreiter1997long} might further significantly improve the prediction quality. Application of techniques from machine translation could be beneficial to introduce multiple classes of connections within semantic networks \cite{vaswani2017attention}. Additionally, combining our approach with unsupervised embedding of scientific literature, as shown in \cite{tshitoyan2019unsupervised} could lead to interesting, dynamic networks.

\textbf{Network Theory} and \textbf{Science of Science} -- Currently, \semnet represents connections between concepts that appear in the scientific literature. This is of course a vast simplification of scientific knowledge, as concepts in natural languages can have a manifold of relations \cite{helbig2006knowledge}. An extension could employ more complex structures for knowledge representation, such as hyper-graphs \cite{shi2015weaving}. The concept list, which represents the nodes of \semnet, can be improved by various different, sophisticated ways for generating of lists of concepts and categories \cite{milojevic2015quantifying, sreenivasan2013quantitative}. The extension to combinations of more than pairs of concepts will lead to more complex knowledge representations. Furthermore, it would be insightful to fold into the semantic network numbers of article citations, which is, at least in the field of science, frequently used as a proxy for scientific impact (see \cite{uzzi2013atypical, martin2013coauthorship, kuhn2014inheritance}, for example). This may enable the prediction of future research directions to be made taking into consideration the highest potential impact, potentially accelerating the evolution of individual scientific knowledge \cite{sinatra2016quantifying,barabasi2018formula}. 

\textbf{Surprisingness} -- In this work, we place pairs of concepts in an abstract high-dimensional space and identify outliers that have unique and potentially valuable properties. It would be interesting to apply more, and different measures of surprisingness. An interesting example is the information-based Bayesian surprise function, which has been introduced in the context of human attention \cite{itti2006bayesian} and successfully applied to the subfield of computational creativity \cite{CompCreat, pinel2015culinary}. In order to achieve further progress, it would be important to further explore and genuinely understand what human scientists consider as \textit{surprising} and \textit{creative}.

\section{Discussion}
We show how to create a semantic network in the field of quantum physics, demonstrate its useage to predict future trends in the field and how it can be used to suggest pairs of concepts, which are not yet investigated jointly, but have distinct network properties. We show how to filter the suggestions for the research agends of an individual scientist. The approach presented here is independent of the discipline of science. As such it can be applied to other fields of research.

This can be interpreted as one potential road towards computer-inspired science, in the following sense: We imagine cases (which we believe is possible) where \semnet produces seeds or inspirations of unusual ideas or directions of thoughts, that a researcher alone might not have thought of. The subsequent, successful interpretation and scientific execution of the suggestions fully remains the task of a creative, human scientist. 


\section*{Acknowledgements}
MK thanks James A. Evans and Sasha Belikov for exciting discussions of metaknowledge research and automation of science, and Jacob G. Foster for a short but influencial conversation at the International Symposium on Science of Science 2016. Furthermore, we would like to acknowledge Nora Tischler, Armin Hochrainer, Robert Fickler, Radek Lapkiewicz, Manuel Erhard and Philipp Haslinger for many interesting discussion on related topics. The authors also thank the APS (American Physical Socienty) for providing access to the database of all published articles in APS journals. The authors thank Xuemei Gu for the illustrations of Figure \ref{fig:CreateNetwork} and \ref{fig:TrainNN}.
This work was supported by the Austrian Academy of Sciences (\"OAW), University of Vienna via the project QUESS and the Austrian Science Fund (FWF) with SFB F40 (FOQUS) and the Erwin Schr\"odinger fellowship No. J4309.

\bibliographystyle{unsrt}
\bibliography{refs}

\clearpage
\newpage

\begin{center}
\textbf{\Large Supplementary Information}
\end{center}

\section{Network theoretical properties used for predictions}
The neural network receives 17 network theoretical properties from \semnet, which we detail here. For a concept $c_i$ and $c_j$, the vector $p_{i,j}=(p_{i,j}^1, p_{i,j}^2, \dots, p_{i,j}^{17})$ corresponds to 17 real valued numbers. \semnet of a specific year $Y$ corresponds to an adjacency matrix, which we denote as $AdjM_Y$. 
\begin{itemize}
\item $p_{i,j}^1=\frac{deg(c_i)}{\max_k(deg(c_k))} \in $[0,1]: normalized degree centrality of first concept $c_i$ (normalized by largest degree centrality in the concept list), i.e. with how many other concept is $c_i$ connected divided by the connection numbers of the concept with most neighboring concepts.
\item $p_{i,j}^2=\frac{deg(c_j)}{\max_k(deg(c_k))} \in $[0,1], normalized degree centrality of second concept $c_j$.
\item $p_{i,j}^3=\frac{\#(c_i)}{\max_k(\#(c_k))} \in $[0,1], number of titles and abstract that concept $c_i$ occures (normalized by number of concept that occures in most articles.
\item $p_{i,j}^4=\frac{\#(c_j)}{\max_k(\#(c_k))} \in $[0,1], number of titles and abstract that concept $c_j$ occures (normalized by number of concept that occures in most articles.
\item $p_{i,j}^5=\frac{AdjM_Y^2}{\sqrt{deg(c_i)\cdot deg(c_j)}} \in $[0,1], ratio of common neighbors, also known as \textit{cosine similarity}.

\item $p_{i,j}^6=\frac{AdjM_Y^2(c_i,c_j)}{max_{k,l} AdjM_Y^2(c_k,c_l)} \in $[0,1], paths of length=2 between $c_i$ and $c_j$ normalized by pair with largest number of paths, at year $Y$.
\item $p_{i,j}^7=\frac{AdjM_{Y-1}^2(c_i,c_j)}{max_{k,l} AdjM_{Y-1}^2(c_k,c_l)} \in $[0,1], paths of length=2 between $c_i$ and $c_j$ normalized by pair with largest number of paths, at year $Y-1$.
\item $p_{i,j}^8=\frac{AdjM_{Y-2}^2(c_i,c_j)}{max_{k,l} AdjM_{Y-2}^2(c_k,c_l)} \in $[0,1], paths of length=2 between $c_i$ and $c_j$ normalized by pair with largest number of paths, at year $Y-2$.

\item $p_{i,j}^9=\frac{AdjM_Y^3(c_i,c_j)}{max_{k,l} AdjM_Y^3(c_k,c_l)} \in $[0,1], paths of length=3 between $c_i$ and $c_j$ normalized by pair with largest number of paths, at year $Y$.
\item $p_{i,j}^{10}=\frac{AdjM_{Y-1}^3(c_i,c_j)}{max_{k,l} AdjM_{Y-1}^3(c_k,c_l)} \in $[0,1], paths of length=3 between $c_i$ and $c_j$ normalized by pair with largest number of paths, at year $Y-1$.
\item $p_{i,j}^{11}=\frac{AdjM_{Y-2}^3(c_i,c_j)}{max_{k,l} AdjM_{Y-2}^3(c_k,c_l)} \in $[0,1], paths of length=3 between $c_i$ and $c_j$ normalized by pair with largest number of paths, at year $Y-2$.

\item $p_{i,j}^{12}=\frac{AdjM_Y^4(c_i,c_j)}{max_{k,l} AdjM_Y^4(c_k,c_l)} \in $[0,1], paths of length=4 between $c_i$ and $c_j$ normalized by pair with largest number of paths, at year $Y$.
\item $p_{i,j}^{13}=\frac{AdjM_{Y-1}^4(c_i,c_j)}{max_{k,l} AdjM_{Y-1}^4(c_k,c_l)} \in $[0,1], paths of length=4 between $c_i$ and $c_j$ normalized by pair with largest number of paths, at year $Y-1$.
\item $p_{i,j}^{14}=\frac{AdjM_{Y-2}^4(c_i,c_j)}{max_{k,l} AdjM_{Y-2}^4(c_k,c_l)} \in $[0,1], paths of length=4 between $c_i$ and $c_j$ normalized by pair with largest number of paths, at year $Y-2$.

\item $p_{i,j}^{15}=distance(c_i,c_j) \in \mathbb{N}$, network distance between $c_i$ and $c_j$.
\item $p_{i,j}^{16}=WeightedDistance(\frac{\sqrt{deg(c_k)\cdot deg(c_l)}}{AdjM_Y(c_k,c_l)}) \in $[0,1], weighted network distance between $c_i$ and $c_j$ (normalized by largest value of all pairs). Intuition: The more connections between certain edges, the easier it to transition from the one to the other.
\item $p_{i,j}^{17}=WeightedDistance(\frac{deg(c_k)\cdot deg(c_l)}{AdjM_Y(c_k,c_l)}) \in $[0,1], different normalized weighted network distance between $c_i$ and $c_j$. Intuition: The more connections between certain edges, the easier it to transition from the one to the other.
\end{itemize}

\section{Future suggestions from \semnet}
Here we show a number of future suggestions with different parameter settings. These pairs of concepts are network-theoretically distinguished, and they couldd be inspirations for the creative, human scientist. The concept list used here is unrestricted, meaning not tailored for a specific scientist's research interest.

\subsection{General Concepts}
\textbf{Unrestricted; Highest predicted values:}
\begin{enumerate}[noitemsep]  
\item hybrid system, classical communication \\cosS: 0.30407, deg: 0.22924, pred: 1
\item back action, classical communication \\cosS: 0.34642, deg: 0.23012, pred: 0.98235
\item spin orbit interaction, quantum sensing \\cosS: 0.31003, deg: 0.23375, pred: 0.95525
\item conformal field theory, classical communication \\cosS: 0.28176, deg: 0.23493, pred: 0.94893
\item spin orbit coupling, quantum sensing \\cosS: 0.33201, deg: 0.25839, pred: 0.94077
\item light matter interaction, classical communication \\cosS: 0.28623, deg: 0.24769, pred: 0.93416
\item classical mechanic, classical communication \\cosS: 0.3182, deg: 0.24956, pred: 0.92603
\item universality, weyl semimetal \\cosS: 0.44731, deg: 0.30365, pred: 0.90986
\item many body physic, classical communication \\cosS: 0.29946, deg: 0.23414, pred: 0.9079
\item propagator, weyl semimetal \\cosS: 0.44141, deg: 0.30493, pred: 0.88731
\end{enumerate}

\textbf{cosS$<$0.15; Highest predicted values:}
\begin{enumerate}[noitemsep]  
\item molecule, stanene \\cosS: 0.14975, deg: 0.38553, pred: 0.87155
\item wave function, stanene \\cosS: 0.14554, deg: 0.41675, pred: 0.85192
\item ground state, laser printing \\cosS: 0.080176, deg: 0.43108, pred: 0.79129
\item laser, stanene \\cosS: 0.14711, deg: 0.39918, pred: 0.73576
\item spin state, rarita schwinger equation \\cosS: 0.10752, deg: 0.25182, pred: 0.73427
\item two level atom, ultracold atom gas \\cosS: 0.14962, deg: 0.20833, pred: 0.71826
\item correlation, laser printing \\cosS: 0.076358, deg: 0.47497, pred: 0.71787
\item optical lattice, electromagnetically induced grating \\cosS: 0.12275, deg: 0.24917, pred: 0.71311
\item polarization, laser printing \\cosS: 0.083372, deg: 0.42666, pred: 0.71008
\item wave function, laser printing \\cosS: 0.082139, deg: 0.41086, pred: 0.70284
\end{enumerate}

\textbf{deg$<$0.05; Highest predicted values:}
\begin{enumerate}[noitemsep]  
\item seesaw mechanism, dark photon \\cosS: 0.42051, deg: 0.046927, pred: 0.52255
\item majoron, tribimaximal mixing \\cosS: 0.43699, deg: 0.026998, pred: 0.4697
\item matrix product operator, multi scale entanglement renormalization ansatz \\cosS: 0.367, deg: 0.044375, pred: 0.45618
\item electron neutrino, tribimaximal mixing \\cosS: 0.32507, deg: 0.047222, pred: 0.45098
\item valleytronic, spin transistor \\cosS: 0.39342, deg: 0.043687, pred: 0.43787
\item fair sampling, bell test experiment \\cosS: 0.38788, deg: 0.018751, pred: 0.4309
\item dark photon, little hierarchy problem \\cosS: 0.4419, deg: 0.026311, pred: 0.4296
\item wiggler, smith purcell effect \\cosS: 0.26696, deg: 0.042411, pred: 0.42564
\item valleytronic, spatial inversion \\cosS: 0.34483, deg: 0.043982, pred: 0.41915
\item quantum key, continuous variable quantum cryptography \\cosS: 0.28986, deg: 0.044375, pred: 0.41585
\end{enumerate}

\textbf{cosS$<$0.15, deg$<$0.05; Highest predicted values:}
\begin{enumerate}[noitemsep]  
\item self pulsing, laser printing \\cosS: 0.13666, deg: 0.028176, pred: 0.22185
\item photosynthesis, laser printing \\cosS: 0.14425, deg: 0.033772, pred: 0.21813
\item neutron capture nucleosynthesis, european spallation source \\cosS: 0.14137, deg: 0.044866, pred: 0.21189
\item apparent violation, eberhard inequality \\cosS: 0.13047, deg: 0.043491, pred: 0.2102
\item copenhagen interpretation, spekkens toy model \\cosS: 0.14746, deg: 0.043393, pred: 0.20579
\item shared entanglement, generalized coherence \\cosS: 0.1419, deg: 0.035833, pred: 0.20522
\item quantum search algorithm, oracle query \\cosS: 0.14003, deg: 0.043197, pred: 0.20485
\item photon counter, photonic orbital angular momentum \\cosS: 0.14217, deg: 0.04192, pred: 0.20478
\item copenhagen interpretation, quasi set theory \\cosS: 0.1326, deg: 0.040349, pred: 0.20417
\item optical amplifier, laser printing \\cosS: 0.14551, deg: 0.042509, pred: 0.20308
\end{enumerate}

\textbf{Unrestricted; Highest predicted values:}
\begin{enumerate}[noitemsep]  
\item hybrid system, classical communication \\cosS: 0.30407, deg: 0.22924, pred: 1
\item back action, classical communication \\cosS: 0.34642, deg: 0.23012, pred: 0.98235
\item spin orbit interaction, quantum sensing \\cosS: 0.31003, deg: 0.23375, pred: 0.95525
\item conformal field theory, classical communication \\cosS: 0.28176, deg: 0.23493, pred: 0.94893
\item spin orbit coupling, quantum sensing \\cosS: 0.33201, deg: 0.25839, pred: 0.94077
\item light matter interaction, classical communication \\cosS: 0.28623, deg: 0.24769, pred: 0.93416
\item classical mechanic, classical communication \\cosS: 0.3182, deg: 0.24956, pred: 0.92603
\item universality, weyl semimetal \\cosS: 0.44731, deg: 0.30365, pred: 0.90986
\item many body physic, classical communication \\cosS: 0.29946, deg: 0.23414, pred: 0.9079
\item propagator, weyl semimetal \\cosS: 0.44141, deg: 0.30493, pred: 0.88731
\end{enumerate}

\textbf{Unrestricted; Lowest predicted values:}
\begin{enumerate}[noitemsep]  
\item transverse mode, pseudogap \\cosS: 0.47207, deg: 0.22227, pred: -1
\item nonlinear regime, pseudogap \\cosS: 0.48811, deg: 0.21971, pred: -0.99384
\item langevin equation, pseudogap \\cosS: 0.48992, deg: 0.24897, pred: -0.99167
\item numerical computation, pseudogap \\cosS: 0.51088, deg: 0.24357, pred: -0.98443
\item diffusion process, pseudogap \\cosS: 0.51135, deg: 0.21971, pred: -0.98135
\item interaction hamiltonian, pseudogap \\cosS: 0.483, deg: 0.24789, pred: -0.98065
\item holography, pseudogap \\cosS: 0.4797, deg: 0.22413, pred: -0.97841
\item many particle system, inelastic neutron scattering \\cosS: 0.46252, deg: 0.20253, pred: -0.97628
\item damping rate, pseudogap \\cosS: 0.49515, deg: 0.21814, pred: -0.97625
\item early universe, pseudogap \\cosS: 0.42681, deg: 0.21716, pred: -0.9754
\end{enumerate}

\textbf{cosS$<$0.15; Lowest predicted values:}
\begin{enumerate}[noitemsep]  
\item laser, large helical device \\cosS: 0.093823, deg: 0.39378, pred: -0.72391
\item distribution, pionium \\cosS: 0.11835, deg: 0.50461, pred: -0.62882
\item laser, diffuse serie \\cosS: 0.10166, deg: 0.39476, pred: -0.61814
\item resolution, moseleys law \\cosS: 0.075495, deg: 0.38111, pred: -0.60875
\item charge, franck hertz experiment \\cosS: 0.085768, deg: 0.44365, pred: -0.55765
\item charge, selected area diffraction \\cosS: 0.10018, deg: 0.44502, pred: -0.55725
\item hamiltonian, zero field nmr \\cosS: 0.14845, deg: 0.4462, pred: -0.55318
\item molecule, atom transition \\cosS: 0.1266, deg: 0.38386, pred: -0.55074
\item electron, atom bose einstein condensate \\cosS: 0.1139, deg: 0.49146, pred: -0.54915
\item electron, ultracold atom gas \\cosS: 0.12406, deg: 0.49224, pred: -0.54876
\end{enumerate}

\textbf{Unrestricted; maximal outlier (cosS, deg, pred):}
\begin{enumerate}[noitemsep]  
\item quantum information, scattering amplitude \\cosS: 0.49361, deg: 0.5376, pred: -0.95502
\item s process, quantum spin \\cosS: 0.59655, deg: 0.48164, pred: -0.95498
\item electrostatic, spin system \\cosS: 0.58982, deg: 0.45376, pred: -0.95086
\item hilbert space, raman scattering \\cosS: 0.48201, deg: 0.47477, pred: -0.95554
\item interference effect, mean field theory \\cosS: 0.58245, deg: 0.38131, pred: -0.95981
\item space time, carbon nanotube \\cosS: 0.51336, deg: 0.42284, pred: -0.95861
\item quantum optic, random phase approximation \\cosS: 0.48734, deg: 0.43, pred: -0.95878
\item quantum information, brillouin zone \\cosS: 0.50927, deg: 0.52562, pred: -0.86694
\item two level system, charge density \\cosS: 0.51577, deg: 0.41223, pred: -0.95105
\item path integral, raman scattering \\cosS: 0.53407, deg: 0.41331, pred: -0.93953
\end{enumerate}

\textbf{Unrestricted; maximal outlier (cosS, deg):}
\begin{enumerate}[noitemsep]  
\item hilbert space, plasma \\cosS: 0.5505, deg: 0.57157, pred: -0.458
\item divergence, quantum computation \\cosS: 0.56671, deg: 0.53466, pred: 0.063652
\item wave packet, free energy \\cosS: 0.60923, deg: 0.50884, pred: -0.55609
\item quantum information, wave number \\cosS: 0.52858, deg: 0.54683, pred: 0.087118
\item atom, yang mills theory \\cosS: 0.39169, deg: 0.58777, pred: 0.019855
\item entangled state, conductivity \\cosS: 0.50832, deg: 0.54752, pred: -0.45379
\item density matrix, domain wall \\cosS: 0.58721, deg: 0.5105, pred: 0.10296
\item qubit, diffusion coefficient \\cosS: 0.52962, deg: 0.53642, pred: 0.11948
\item entanglement, vector potential \\cosS: 0.50925, deg: 0.54271, pred: 0.11929
\item decoherence, electromagnetic wave \\cosS: 0.5603, deg: 0.51885, pred: 0.095746
\end{enumerate}

\section{Network theoretical properties used for predictions}
The neural network receives 17 network theoretical properties from \semnet, which we detail here. For a concept $c_i$ and $c_j$, the vector $p_{i,j}=(p_{i,j}^1, p_{i,j}^2, \dots, p_{i,j}^{17})$ corresponds to 17 real valued numbers. \semnet of a specific year $Y$ corresponds to an adjacency matrix, which we denote as $AdjM_Y$. 
\begin{itemize}
\item $p_{i,j}^1=\frac{deg(c_i)}{\max_k(deg(c_k))} \in $[0,1]: normalized degree centrality of first concept $c_i$ (normalized by largest degree centrality in the concept list), i.e. with how many other concept is $c_i$ connected divided by the connection numbers of the concept with most neighboring concepts.
\item $p_{i,j}^2=\frac{deg(c_j)}{\max_k(deg(c_k))} \in $[0,1], normalized degree centrality of second concept $c_j$.
\item $p_{i,j}^3=\frac{\#(c_i)}{\max_k(\#(c_k))} \in $[0,1], number of titles and abstract that concept $c_i$ occures (normalized by number of concept that occures in most articles.
\item $p_{i,j}^4=\frac{\#(c_j)}{\max_k(\#(c_k))} \in $[0,1], number of titles and abstract that concept $c_j$ occures (normalized by number of concept that occures in most articles.
\item $p_{i,j}^5=\frac{AdjM_Y^2}{\sqrt{deg(c_i)\cdot deg(c_j)}} \in $[0,1], ratio of common neighbors, also known as \textit{cosine similarity}.

\item $p_{i,j}^6=\frac{AdjM_Y^2(c_i,c_j)}{max_{k,l} AdjM_Y^2(c_k,c_l)} \in $[0,1], paths of length=2 between $c_i$ and $c_j$ normalized by pair with largest number of paths, at year $Y$.
\item $p_{i,j}^7=\frac{AdjM_{Y-1}^2(c_i,c_j)}{max_{k,l} AdjM_{Y-1}^2(c_k,c_l)} \in $[0,1], paths of length=2 between $c_i$ and $c_j$ normalized by pair with largest number of paths, at year $Y-1$.
\item $p_{i,j}^8=\frac{AdjM_{Y-2}^2(c_i,c_j)}{max_{k,l} AdjM_{Y-2}^2(c_k,c_l)} \in $[0,1], paths of length=2 between $c_i$ and $c_j$ normalized by pair with largest number of paths, at year $Y-2$.

\item $p_{i,j}^9=\frac{AdjM_Y^3(c_i,c_j)}{max_{k,l} AdjM_Y^3(c_k,c_l)} \in $[0,1], paths of length=3 between $c_i$ and $c_j$ normalized by pair with largest number of paths, at year $Y$.
\item $p_{i,j}^{10}=\frac{AdjM_{Y-1}^3(c_i,c_j)}{max_{k,l} AdjM_{Y-1}^3(c_k,c_l)} \in $[0,1], paths of length=3 between $c_i$ and $c_j$ normalized by pair with largest number of paths, at year $Y-1$.
\item $p_{i,j}^{11}=\frac{AdjM_{Y-2}^3(c_i,c_j)}{max_{k,l} AdjM_{Y-2}^3(c_k,c_l)} \in $[0,1], paths of length=3 between $c_i$ and $c_j$ normalized by pair with largest number of paths, at year $Y-2$.

\item $p_{i,j}^{12}=\frac{AdjM_Y^4(c_i,c_j)}{max_{k,l} AdjM_Y^4(c_k,c_l)} \in $[0,1], paths of length=4 between $c_i$ and $c_j$ normalized by pair with largest number of paths, at year $Y$.
\item $p_{i,j}^{13}=\frac{AdjM_{Y-1}^4(c_i,c_j)}{max_{k,l} AdjM_{Y-1}^4(c_k,c_l)} \in $[0,1], paths of length=4 between $c_i$ and $c_j$ normalized by pair with largest number of paths, at year $Y-1$.
\item $p_{i,j}^{14}=\frac{AdjM_{Y-2}^4(c_i,c_j)}{max_{k,l} AdjM_{Y-2}^4(c_k,c_l)} \in $[0,1], paths of length=4 between $c_i$ and $c_j$ normalized by pair with largest number of paths, at year $Y-2$.

\item $p_{i,j}^{15}=distance(c_i,c_j) \in \mathbb{N}$, network distance between $c_i$ and $c_j$.
\item $p_{i,j}^{16}=WeightedDistance(\frac{\sqrt{deg(c_k)\cdot deg(c_l)}}{AdjM_Y(c_k,c_l)}) \in $[0,1], weighted network distance between $c_i$ and $c_j$ (normalized by largest value of all pairs). Intuition: The more connections between certain edges, the easier it to transition from the one to the other.
\item $p_{i,j}^{17}=WeightedDistance(\frac{deg(c_k)\cdot deg(c_l)}{AdjM_Y(c_k,c_l)}) \in $[0,1], different normalized weighted network distance between $c_i$ and $c_j$. Intuition: The more connections between certain edges, the easier it to transition from the one to the other.
\end{itemize}

\section{Future suggestions from \semnet}
Here we show a number of future suggestions with different parameter settings. These pairs of concepts are network-theoretically distinguished, and they couldd be inspirations for the creative, human scientist. The concept list used here is unrestricted, meaning not tailored for a specific scientist's research interest.

\subsection{General Concepts}
\textbf{Unrestricted; Highest predicted values:}
\begin{enumerate}[noitemsep]  
\item hybrid system, classical communication \\cosS: 0.30407, deg: 0.22924, pred: 1
\item back action, classical communication \\cosS: 0.34642, deg: 0.23012, pred: 0.98235
\item spin orbit interaction, quantum sensing \\cosS: 0.31003, deg: 0.23375, pred: 0.95525
\item conformal field theory, classical communication \\cosS: 0.28176, deg: 0.23493, pred: 0.94893
\item spin orbit coupling, quantum sensing \\cosS: 0.33201, deg: 0.25839, pred: 0.94077
\item light matter interaction, classical communication \\cosS: 0.28623, deg: 0.24769, pred: 0.93416
\item classical mechanic, classical communication \\cosS: 0.3182, deg: 0.24956, pred: 0.92603
\item universality, weyl semimetal \\cosS: 0.44731, deg: 0.30365, pred: 0.90986
\item many body physic, classical communication \\cosS: 0.29946, deg: 0.23414, pred: 0.9079
\item propagator, weyl semimetal \\cosS: 0.44141, deg: 0.30493, pred: 0.88731
\end{enumerate}

\textbf{cosS$<$0.15; Highest predicted values:}
\begin{enumerate}[noitemsep]  
\item molecule, stanene \\cosS: 0.14975, deg: 0.38553, pred: 0.87155
\item wave function, stanene \\cosS: 0.14554, deg: 0.41675, pred: 0.85192
\item ground state, laser printing \\cosS: 0.080176, deg: 0.43108, pred: 0.79129
\item laser, stanene \\cosS: 0.14711, deg: 0.39918, pred: 0.73576
\item spin state, rarita schwinger equation \\cosS: 0.10752, deg: 0.25182, pred: 0.73427
\item two level atom, ultracold atom gas \\cosS: 0.14962, deg: 0.20833, pred: 0.71826
\item correlation, laser printing \\cosS: 0.076358, deg: 0.47497, pred: 0.71787
\item optical lattice, electromagnetically induced grating \\cosS: 0.12275, deg: 0.24917, pred: 0.71311
\item polarization, laser printing \\cosS: 0.083372, deg: 0.42666, pred: 0.71008
\item wave function, laser printing \\cosS: 0.082139, deg: 0.41086, pred: 0.70284
\end{enumerate}

\textbf{deg$<$0.05; Highest predicted values:}
\begin{enumerate}[noitemsep]  
\item seesaw mechanism, dark photon \\cosS: 0.42051, deg: 0.046927, pred: 0.52255
\item majoron, tribimaximal mixing \\cosS: 0.43699, deg: 0.026998, pred: 0.4697
\item matrix product operator, multi scale entanglement renormalization ansatz \\cosS: 0.367, deg: 0.044375, pred: 0.45618
\item electron neutrino, tribimaximal mixing \\cosS: 0.32507, deg: 0.047222, pred: 0.45098
\item valleytronic, spin transistor \\cosS: 0.39342, deg: 0.043687, pred: 0.43787
\item fair sampling, bell test experiment \\cosS: 0.38788, deg: 0.018751, pred: 0.4309
\item dark photon, little hierarchy problem \\cosS: 0.4419, deg: 0.026311, pred: 0.4296
\item wiggler, smith purcell effect \\cosS: 0.26696, deg: 0.042411, pred: 0.42564
\item valleytronic, spatial inversion \\cosS: 0.34483, deg: 0.043982, pred: 0.41915
\item quantum key, continuous variable quantum cryptography \\cosS: 0.28986, deg: 0.044375, pred: 0.41585
\end{enumerate}

\textbf{cosS$<$0.15, deg$<$0.05; Highest predicted values:}
\begin{enumerate}[noitemsep]  
\item self pulsing, laser printing \\cosS: 0.13666, deg: 0.028176, pred: 0.22185
\item photosynthesis, laser printing \\cosS: 0.14425, deg: 0.033772, pred: 0.21813
\item neutron capture nucleosynthesis, european spallation source \\cosS: 0.14137, deg: 0.044866, pred: 0.21189
\item apparent violation, eberhard inequality \\cosS: 0.13047, deg: 0.043491, pred: 0.2102
\item copenhagen interpretation, spekkens toy model \\cosS: 0.14746, deg: 0.043393, pred: 0.20579
\item shared entanglement, generalized coherence \\cosS: 0.1419, deg: 0.035833, pred: 0.20522
\item quantum search algorithm, oracle query \\cosS: 0.14003, deg: 0.043197, pred: 0.20485
\item photon counter, photonic orbital angular momentum \\cosS: 0.14217, deg: 0.04192, pred: 0.20478
\item copenhagen interpretation, quasi set theory \\cosS: 0.1326, deg: 0.040349, pred: 0.20417
\item optical amplifier, laser printing \\cosS: 0.14551, deg: 0.042509, pred: 0.20308
\end{enumerate}

\textbf{Unrestricted; Highest predicted values:}
\begin{enumerate}[noitemsep]  
\item hybrid system, classical communication \\cosS: 0.30407, deg: 0.22924, pred: 1
\item back action, classical communication \\cosS: 0.34642, deg: 0.23012, pred: 0.98235
\item spin orbit interaction, quantum sensing \\cosS: 0.31003, deg: 0.23375, pred: 0.95525
\item conformal field theory, classical communication \\cosS: 0.28176, deg: 0.23493, pred: 0.94893
\item spin orbit coupling, quantum sensing \\cosS: 0.33201, deg: 0.25839, pred: 0.94077
\item light matter interaction, classical communication \\cosS: 0.28623, deg: 0.24769, pred: 0.93416
\item classical mechanic, classical communication \\cosS: 0.3182, deg: 0.24956, pred: 0.92603
\item universality, weyl semimetal \\cosS: 0.44731, deg: 0.30365, pred: 0.90986
\item many body physic, classical communication \\cosS: 0.29946, deg: 0.23414, pred: 0.9079
\item propagator, weyl semimetal \\cosS: 0.44141, deg: 0.30493, pred: 0.88731
\end{enumerate}

\textbf{Unrestricted; Lowest predicted values:}
\begin{enumerate}[noitemsep]  
\item transverse mode, pseudogap \\cosS: 0.47207, deg: 0.22227, pred: -1
\item nonlinear regime, pseudogap \\cosS: 0.48811, deg: 0.21971, pred: -0.99384
\item langevin equation, pseudogap \\cosS: 0.48992, deg: 0.24897, pred: -0.99167
\item numerical computation, pseudogap \\cosS: 0.51088, deg: 0.24357, pred: -0.98443
\item diffusion process, pseudogap \\cosS: 0.51135, deg: 0.21971, pred: -0.98135
\item interaction hamiltonian, pseudogap \\cosS: 0.483, deg: 0.24789, pred: -0.98065
\item holography, pseudogap \\cosS: 0.4797, deg: 0.22413, pred: -0.97841
\item many particle system, inelastic neutron scattering \\cosS: 0.46252, deg: 0.20253, pred: -0.97628
\item damping rate, pseudogap \\cosS: 0.49515, deg: 0.21814, pred: -0.97625
\item early universe, pseudogap \\cosS: 0.42681, deg: 0.21716, pred: -0.9754
\end{enumerate}

\textbf{cosS$<$0.15; Lowest predicted values:}
\begin{enumerate}[noitemsep]  
\item laser, large helical device \\cosS: 0.093823, deg: 0.39378, pred: -0.72391
\item distribution, pionium \\cosS: 0.11835, deg: 0.50461, pred: -0.62882
\item laser, diffuse serie \\cosS: 0.10166, deg: 0.39476, pred: -0.61814
\item resolution, moseleys law \\cosS: 0.075495, deg: 0.38111, pred: -0.60875
\item charge, franck hertz experiment \\cosS: 0.085768, deg: 0.44365, pred: -0.55765
\item charge, selected area diffraction \\cosS: 0.10018, deg: 0.44502, pred: -0.55725
\item hamiltonian, zero field nmr \\cosS: 0.14845, deg: 0.4462, pred: -0.55318
\item molecule, atom transition \\cosS: 0.1266, deg: 0.38386, pred: -0.55074
\item electron, atom bose einstein condensate \\cosS: 0.1139, deg: 0.49146, pred: -0.54915
\item electron, ultracold atom gas \\cosS: 0.12406, deg: 0.49224, pred: -0.54876
\end{enumerate}

\textbf{Unrestricted; maximal outlier (cosS, deg, pred):}
\begin{enumerate}[noitemsep]  
\item quantum information, scattering amplitude \\cosS: 0.49361, deg: 0.5376, pred: -0.95502
\item s process, quantum spin \\cosS: 0.59655, deg: 0.48164, pred: -0.95498
\item electrostatic, spin system \\cosS: 0.58982, deg: 0.45376, pred: -0.95086
\item hilbert space, raman scattering \\cosS: 0.48201, deg: 0.47477, pred: -0.95554
\item interference effect, mean field theory \\cosS: 0.58245, deg: 0.38131, pred: -0.95981
\item space time, carbon nanotube \\cosS: 0.51336, deg: 0.42284, pred: -0.95861
\item quantum optic, random phase approximation \\cosS: 0.48734, deg: 0.43, pred: -0.95878
\item quantum information, brillouin zone \\cosS: 0.50927, deg: 0.52562, pred: -0.86694
\item two level system, charge density \\cosS: 0.51577, deg: 0.41223, pred: -0.95105
\item path integral, raman scattering \\cosS: 0.53407, deg: 0.41331, pred: -0.93953
\end{enumerate}

\textbf{Unrestricted; maximal outlier (cosS, deg):}
\begin{enumerate}[noitemsep]  
\item hilbert space, plasma \\cosS: 0.5505, deg: 0.57157, pred: -0.458
\item divergence, quantum computation \\cosS: 0.56671, deg: 0.53466, pred: 0.063652
\item wave packet, free energy \\cosS: 0.60923, deg: 0.50884, pred: -0.55609
\item quantum information, wave number \\cosS: 0.52858, deg: 0.54683, pred: 0.087118
\item atom, yang mills theory \\cosS: 0.39169, deg: 0.58777, pred: 0.019855
\item entangled state, conductivity \\cosS: 0.50832, deg: 0.54752, pred: -0.45379
\item density matrix, domain wall \\cosS: 0.58721, deg: 0.5105, pred: 0.10296
\item qubit, diffusion coefficient \\cosS: 0.52962, deg: 0.53642, pred: 0.11948
\item entanglement, vector potential \\cosS: 0.50925, deg: 0.54271, pred: 0.11929
\item decoherence, electromagnetic wave \\cosS: 0.5603, deg: 0.51885, pred: 0.095746
\end{enumerate}

\end{document}